\documentclass{optica-article}
\usepackage{makecell}

\journal{opticajournal} 

\articletype{Research Article}

\usepackage{lineno}
\usepackage{booktabs} 
\usepackage{array}
\usepackage{ragged2e}   

\begin{document}

\title{Compact and stable source of polarization-entangled photon-pairs based on a folded linear displacement interferometer}

\author{Sarah E. McCarthy\authormark{1,2,*}, Ali Anwar\authormark{2}, Daniel K. L. Oi\authormark{1} and Loyd J. McKnight\authormark{1,2}}

\address{\authormark{1}SUPA Department of Physics, University of Strathclyde, Glasgow G4 0NG, UK\\
\authormark{2}Fraunhofer Center for Applied Photonics, Glasgow G1 1RD, UK }

\email{\authormark{*}sarah.mccarthy@fraunhofer.co.uk, sarah.mccarthy@strath.ac.uk} 


\begin{abstract*} 
The realization of quantum networks requires the development of robust low size, weight and power (SWaP) systems suitable for operation under harsh environments in remote and mobile nodes such as satellites. We present a source of polarization-entangled photon-pairs in a folded linear displacement interferometer based on spontaneous parametric down conversion using a Type-0 periodically poled potassium titanyl phosphate crystal. Featuring a compact and stable double-pass geometry using a corner-cube retroreflector, the source has a detected pair rate of 2.5~M~pairs/s/mW with a Bell state fidelity of 94.1\% $\pm$ 2.1\%. The qualities and demonstrated performance of the source make it suitable for deployment in entanglement-based quantum networks.

\end{abstract*}

\section{Introduction}

Quantum entanglement is a key resource in many quantum technologies, providing enhancements in imaging and sensing\cite{Pirandola_2018,camphausen23}. Most notably it is the basis of many quantum communication schemes, such as ensuring the security of quantum key distribution (QKD) protocols E91~\cite{ekert91} and BBM92~\cite{BBM92}, with broader entanglement distribution predicted to form the basis of distributed quantum computation\cite{Caleffi_2024} and quantum internet architectures~\cite{pant2017}. For longer range communications, satellite-to-ground links offer a lower loss alternative to fiber networks due to a more favourable loss scaling~\cite{Liao_2017}. Increasingly, small satellite platforms are being used in the deployment of experimental payloads containing different quantum systems such as entangled photon-pair sources, as they provide a cheaper and faster route to in-orbit demonstrations~\cite{oi2017, Oi_apr2017}. These sources need to maintain performance in a highly restricted footprint and under harsh environmental conditions. As a result of these challenges, there have been only a few experimental demonstrations of entanglement generation in space~\cite{Yin2017, Lu2022, villar2020}.

Despite advances in alternative approaches, such as deterministic quantum emitters~\cite{schimpf21} and spontaneous four-wave mixing~\cite{wang24}, spontaneous parametric down-conversion (SPDC) remains a standard method for generating entangled photon-pairs. In this process, a pump photon passing through a second-order nonlinear optical medium generates a photon-pair which is correlated in different degrees of freedom, such as polarization~\cite{Li_2015}, frequency~\cite{Olislager2009}, or time-bin~\cite{Kwon13}, according to the conservation of energy and momentum.

Although there is a large variety of entangled photon-pair source designs~\cite{Anwar2021}, there are only a few that are practical for real-world applications. Some sources with conventional interferometer-based designs, such as the Sagnac geometry~\cite{steinlechner2014}, can be designed and built to be stable with high pair production rates. However, these sources require a larger footprint and active alignment techniques, which make them less suitable for remote deployment in quantum networks. Sources with linear geometries have demonstrated stability against harsh environments and therefore are attractive for future in-orbit demonstrations of entanglement generation and distribution~\cite{Anwar22}.

A source based on two birefringent phase-matched crystals with parallel optical axes~\cite{villar2018} on a CubeSat platform demonstrated entanglement in space~\cite{villar2020}. Sources with Linear Displacement Interferometer (LDI) designs have been shown to achieve the pair production rates required to enable satellite-mediated quantum communication channels~\cite{perumangatt2021realizing}. In a single LDI geometry~\cite{Lohrmann2020}, separate beam displacer (BD) crystals are used to split the pump into two parallel beams and combine the down-converted photon-pairs to generate the entangled state. In a double LDI geometry~\cite{horn2019auto, fazili2024simple}, two BD crystals are used to split the pump and another two are used to combine the polarization components. Additionally, such a configuration requires two identical quasi-phase matching (QPM) crystals for pair generation, which are prone to fabrication errors. 

In this paper, a novel configuration for the generation of polarization-entangled photon-pairs based on a folded linear displacement interferometer (FLDI) is demonstrated. The FLDI source uses a double-pass configuration where the pump, after reflection from a corner cube retroreflector (CCR), is passed through the same QPM crystal, doubling the length of the interaction between the pump and the non-linear material and therefore increasing the probability of the down conversion process occurring. Unlike in a single LDI geometry, a single BD crystal is used to split and combine the beams for entanglement generation. This folded geometry is stable due to the insensitivity of the CCR to tip-tilt misalignment. The compactness and robustness of the FLDI design make it a suitable candidate for sources within a future quantum satellite network. The following sections present the concept of the entanglement generation, the experimental set-up used, and, finally, a discussion on the performance of the source demonstrated.

\section{Source Concept}
\label{sec:sourceconcept}
\begin{figure}[b]
\centering\includegraphics[width=0.7\linewidth]{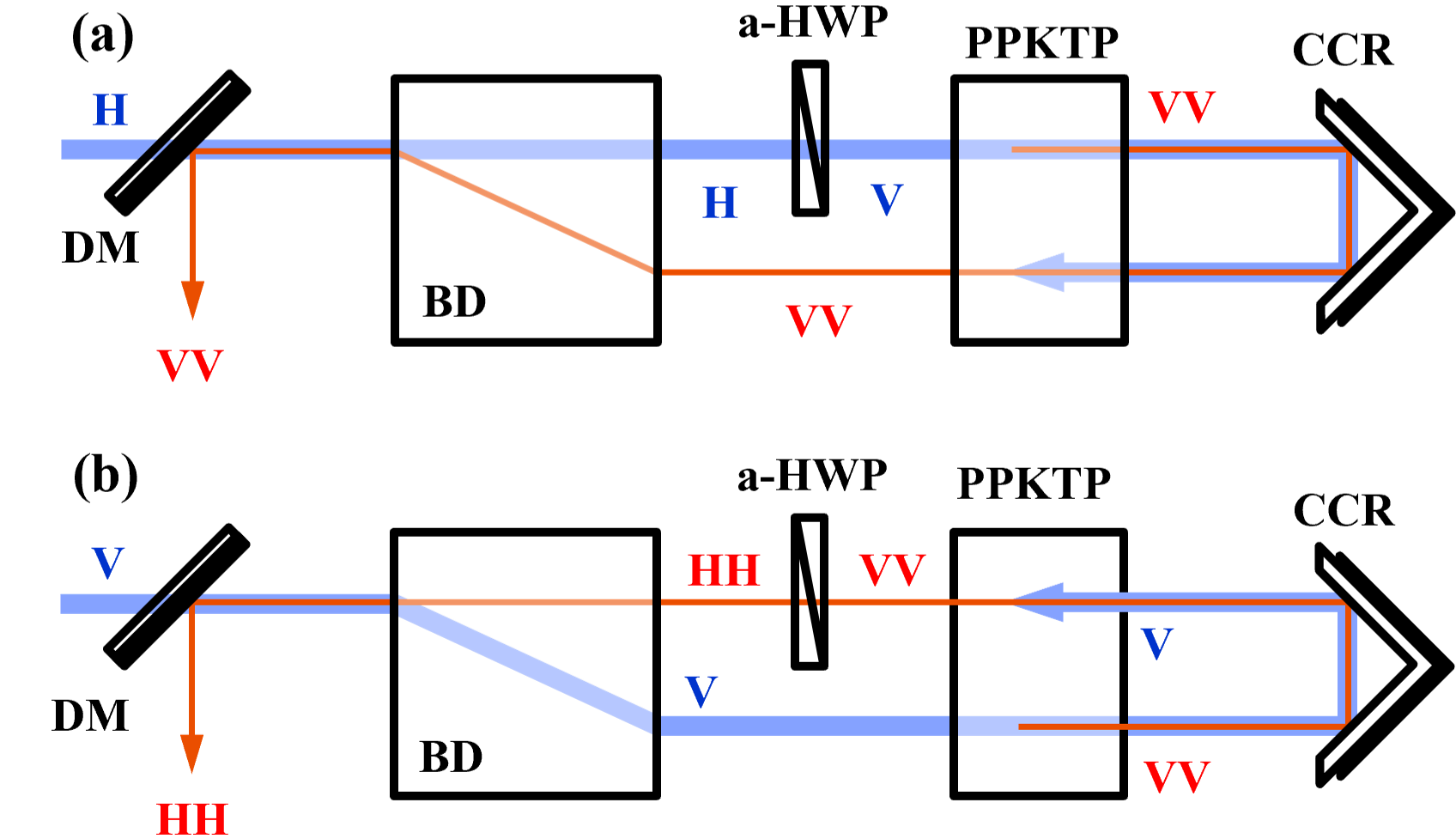}
\caption{FLDI configuration. (a) and (b) show the method of generating polarization entangled state with beam paths of the horizontal and the vertical components of the pump, respectively. The polarization components of the pump counter-propagate around the same path so that vertically polarized pump photons are down converted to VV photon pairs, in both cases, with the a-HWP acting on the pump in (a) and on the SPDC photons in (b).}
\label{fig:entanglement}
\end{figure}

The method for generating the entangled state in the source is illustrated in Fig. \ref{fig:system}~(a) \& (b). In the clockwise direction shown in Fig.~\ref{fig:entanglement}~(a), the H-polarized component of the diagonally polarized input pump beam passes through a beam displacer (BD) without deviation. An achromatic half-wave plate (a-HWP), for both the pump and SPDC wavelengths, oriented at 45 degrees flips the incoming polarization of the beam to V, which passes through the Type-0 periodically poled potassium titanyl phosphate (PPKTP), producing pairs of photons with the same polarization (VV). Both the residual pump and the photon-pairs are then retro-reflected at the mirror (CCR). When returning, the pump passes through the crystal again, allowing the pump beam another opportunity to be downconverted to produce a pair of photons in VV polarization. Photon-pairs generated on either pass travel in the same path and get displaced after passing through the BD, closing the loop. 

Similarly, in the counterclockwise direction shown in Fig.~\ref{fig:entanglement}~(b), the V-polarized component of the pump deviates from the BD and produces photon-pairs in its initial and reflected paths through the crystal. The pairs after the double pass convert to HH-polarization after passing through the a-HWP, and come out of the BD without deviation. The counter-propagating nature of the interferometer is analogous to the Sagnac interferometer, though this offers a much smaller footprint and the additional benefits of the double-pass and the CCR stability. The indistinguishability between the HH and VV pairs generated when clockwise and counterclockwise paths recombine at the BD, giving a maximally entangled state:
\begin{equation}
    |\Phi^\phi\rangle = \frac{1}{\sqrt{2}} (|HH\rangle + e^{i\phi}|VV\rangle)
\end{equation}
Here, $\phi$ is the phase between the $HH$ and $VV$ components of the superposition, controlled by the phase of the input state, $\phi = 0, \pi$ gives the Bell states $|\Phi^+\rangle$ and $|\Phi^-\rangle$, respectively. 

In this way, the folded interferometer operates like a Sagnac interferometer as the two elements of the entangled state are generated on counter-propagating paths around the interferometer before being recombined to give the superposition. However, the FLDI source does this in a much smaller footprint and with the additional probability of pair generation given by the double pass through the non-linear medium.

\section{Experiment}

\begin{figure}[t]
\centering\includegraphics[width=0.9\linewidth]{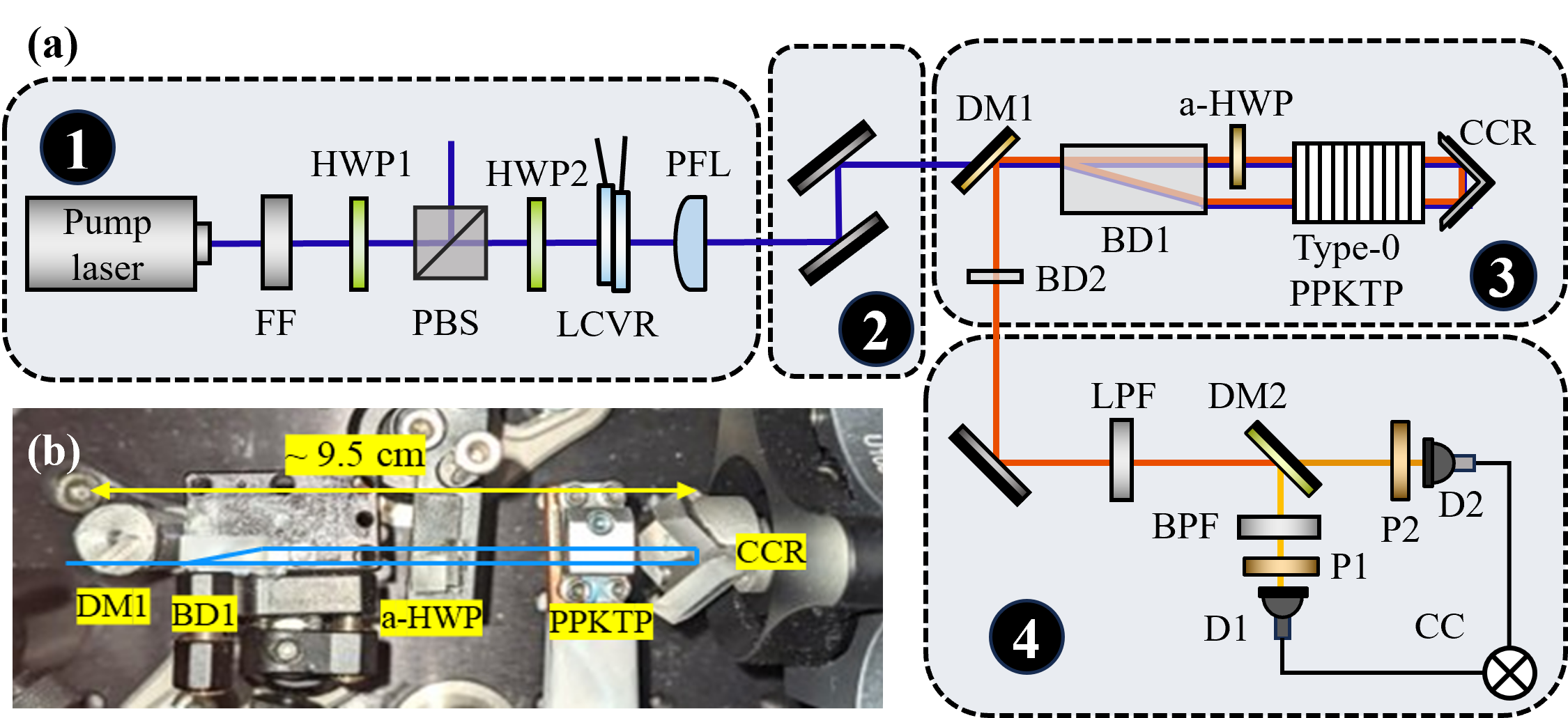}
\caption{FLDI configuration.(a) Experimental schematic of the FLDI source in four sections: (1) pump preparation, (2) beam steering, (3) entanglement generation and (4) detection. FF: Fluorescent filter, HWP: half-wave plate, PBS: polarizing beam splitter, PFL: pump focusing lens, DM: dichroic mirror, BD: beam displacer, CCR: cornercube retroreflector, LPF: long pass filter, BPF: bandpass filter, P: linear polarizer, D: Single photon detectors, CC: coincidence counter. (b) Photograph of section (3), the entanglement generation. Featuring DM1,  BD1, a-HWP, PPKTP and CCR, the interferometer has a of length $\sim 9.5$~cm.}
\label{fig:system}
\end{figure}

The experimental setup for the FLDI source (Fig.~\ref{fig:system}~(a)) is divided into four sections: (1) pump power and polarization control, (2) beam steering, (3) entanglement generation and, finally, (4) detection. A single-mode, broadband continuous-wave laser at 405~nm with a spectral bandwidth of 0.7~nm (Thorlabs LP405C1) pumps a Type-0 PPKTP (Raicol Crystals) of dimensions $1 \times 2 \times 10$~mm and poling-period, $\Lambda$ = 3.425~$\mu$m. A fluorescent filter (FF) is used to filter unwanted emission from the pump. A combination of a half-wave plate (HWP1) and a polarizing beam splitter (PBS) is used for optical control of pump power. A second half-wave plate (HWP2) is used to adjust the polarization of the beam. A liquid crystal variable retarder (LCVR) controls the phase between the two photon-pairs generated by adjusting the H and V components of the pump beam.

The pump light is focused into the PPKTP by a planoconvex lens (pump focussing lens, PFL) of focal length 200~mm. The beam steering consists of two mirrors that direct the pump into the PPKTP.
A beam displacer (BD) crystal ($\alpha$-BBO, AG Optics) of length 13~mm is used to split the incoming beam into two parallel beams with orthogonal polarizations to each other separated by approximately 1 mm. The orthogonal polarization components then follow the entanglement generation scheme described in Section \ref{sec:sourceconcept}.
This results in a compact interferometer with an overall length of $\sim 9.5$~cm (Fig.~\ref{fig:system} (b)) whilst still allowing access for measurement of the power incident on the PPKTP. Thus, there is scope for further size reduction with optimized mounting of the components, for example, in a packaged system.

The SPDC photons are filtered from the pump using a dichroic mirror (DM1). A long-pass filter (LPF) removes any residual pump. Due to birefringence, there will be a slight difference in the length of the BD required to displace the pump and recombine the downconverted photons. Therefore, to ensure complete overlap of the of the polarization components in the output state, a second BD, of length 0.73~mm, is placed before the LPF.
The photon-pairs at non-degenerate wavelengths (Signal: 780~nm, Idler: 842~nm) are separated using a dichroic mirror (DM2). Two linear polarizers, P1 \& P2, are mounted on the signal and idler arms for polarization characterization measurements, respectively. The photons are coupled into a single-mode fiber for detection on single photon avalanche diodes (SPAD, D1 \& D2, SPCM-AQRH-14-FC, Excelitas) and the correlated pairs are recorded using a coincidence counter (CC, TimeTagger Ultra, Swabian Instruments).

An uncoated solid or hollow CCR typically changes the polarization of the beam changes after reflection, inducing some ellipticity in incoming linearly polarized light~\cite{Liu1997, Li2024}. A metal-coated CCR minimises this effect, maintaining the polarization of the input beam, hence a silver-coated hollow CCR (Thorlabs HRR202-P01) was used in the source. A characterization of the polarization shifts from different CCRs is given in Appendix~\ref{appendix}.

\section{Results}
For SPDC utilizing a QPM medium such as PPKTP, the phase-matching condition is~\cite{BOYD202065},
\begin{equation}\label{eqn:QPM}
\Delta k = \frac{2\pi n(\lambda_p,T)}{\lambda_p} - \frac{2\pi n(\lambda_s,T)}{\lambda_s}-\frac{2\pi n(\lambda_i,T)}{\lambda_i} -\frac{2\pi}{\Lambda} \approx 0 ,
\end{equation}
\begin{figure}[h]
\centering\includegraphics[width=0.9\linewidth]{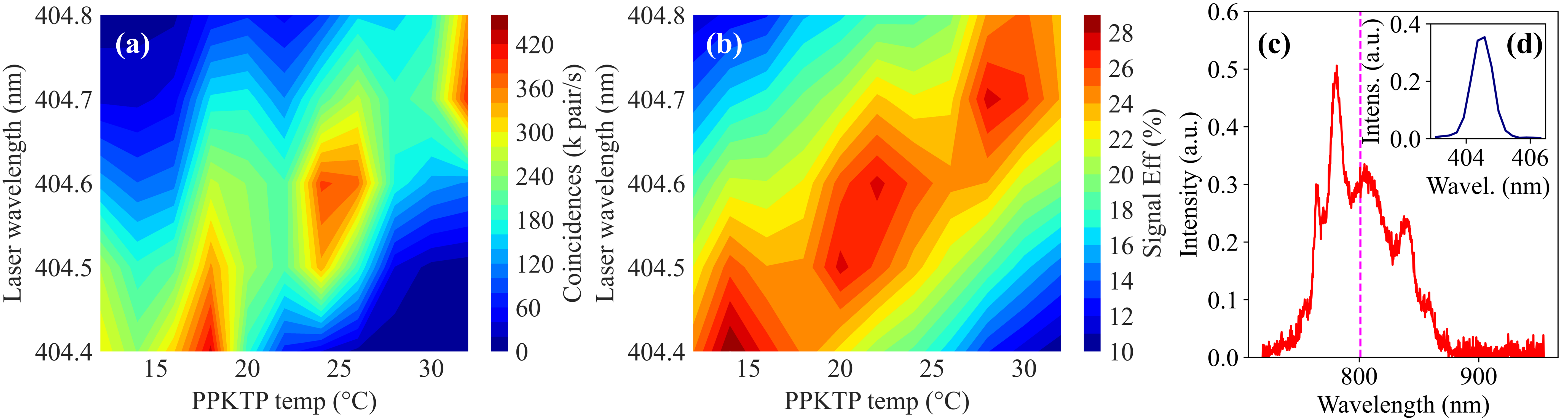}
\caption{Quasi-phase matching. Performance of the source with the wavelength of the pump light, tuned by the laser temperature, and the temperature PPKTP crystal, showing the effect on the (a) coincidence rate and the (b) signal heralding efficiencies, with 10~nm spectral filtering on the signal arm. The laser diode driving current was 50~mA, pumping with 3~mW. Under the same pumping conditions the spectrum of the SPDC and pump is given in (c) and (d) respectively. This was done at a PPKTP temperature of $25.5^\circ$ C, with an integration time of 5~s for the SPDC measurement. The fuchsia line indicates the cut-off wavelength of DM2.}
\label{fig:phasematching}
\end{figure}
where $n$ is the temperature-dependent refractive index and $\lambda_p$, $\lambda_s$ and $\lambda_i$, are the pump, signal and idler wavelengths, respectively.
$T$ and $\Lambda$ are the respective temperature and poling period of the crystal. The parameters $\lambda_p$ and $T$ are usually tuned to optimize the source. Varying the temperature of the laser diode changes $\lambda_p$, which in-turn changes the phase-matching wavelengths of the photon-pairs. The phase-matched generation of photon-pairs is then achieved by temperature tuning of the crystal.
Fig.~\ref{fig:phasematching} shows 2-D plots of the dependence of detected pair rate and the heralding efficiency of signal with pump wavelength and crystal temperature. To optimize the source output, the PPKTP temperature was varied to find values to meet the QPM condition given in Eq.~\ref{eqn:QPM}.

Multiple peaks are observed in both contour plots. Due to the trade-off between the pair rate and the heralding efficiency in the SPDC process, the spectral conditions for optimal pair rate do not necessarily overlap with those of the heralding efficiency. The focusing of the pump and SPDC modes in the crystal is quantified by a dimensionless parameter $\xi$, defined as~\cite{dixon14}
\begin{equation}
    \xi_j=\frac{L}{k_jw_j^2} \quad (j=p, s, i)
\end{equation}
where $L$ is the length of the crystal, $k_j$ and $w_j$ are the respective wavevectors and beam waists of pump, signal and idler modes. Change in the spectra of pump and SPDC affects the focal parameter $\xi$, showing visible peaks and troughs in the contour plots. For a given fixed waists for all the three modes, Stronger focusing of the pump in the SPDC crystal enhances mode overlap leading to an optimal pair production with reduced heralding efficiencies. For a chosen optimal condition from (a) and (b), the measured spectra of the SPDC photons and the pump beam are shown in Fig.~\ref{eqn:QPM}~(c) and (d). The laser was heated to $40^\circ$ C, corresponding to a central wavelength of 404.6~nm. The SPDC spectrum is broadband with multiple peaks observed
near degenerate (810~nm) and non-degenerate regimes.
The largest peak is the signal at 781.3~nm, with smaller peak for the idler at 845.6~nm. Placing a 10~nm bandpass filter centred at 780~nm ensures detection of signal photon at 780 nm and corresponding phase matched idler photon.

\begin{figure}[h]
\centering\includegraphics[width=0.85\linewidth]{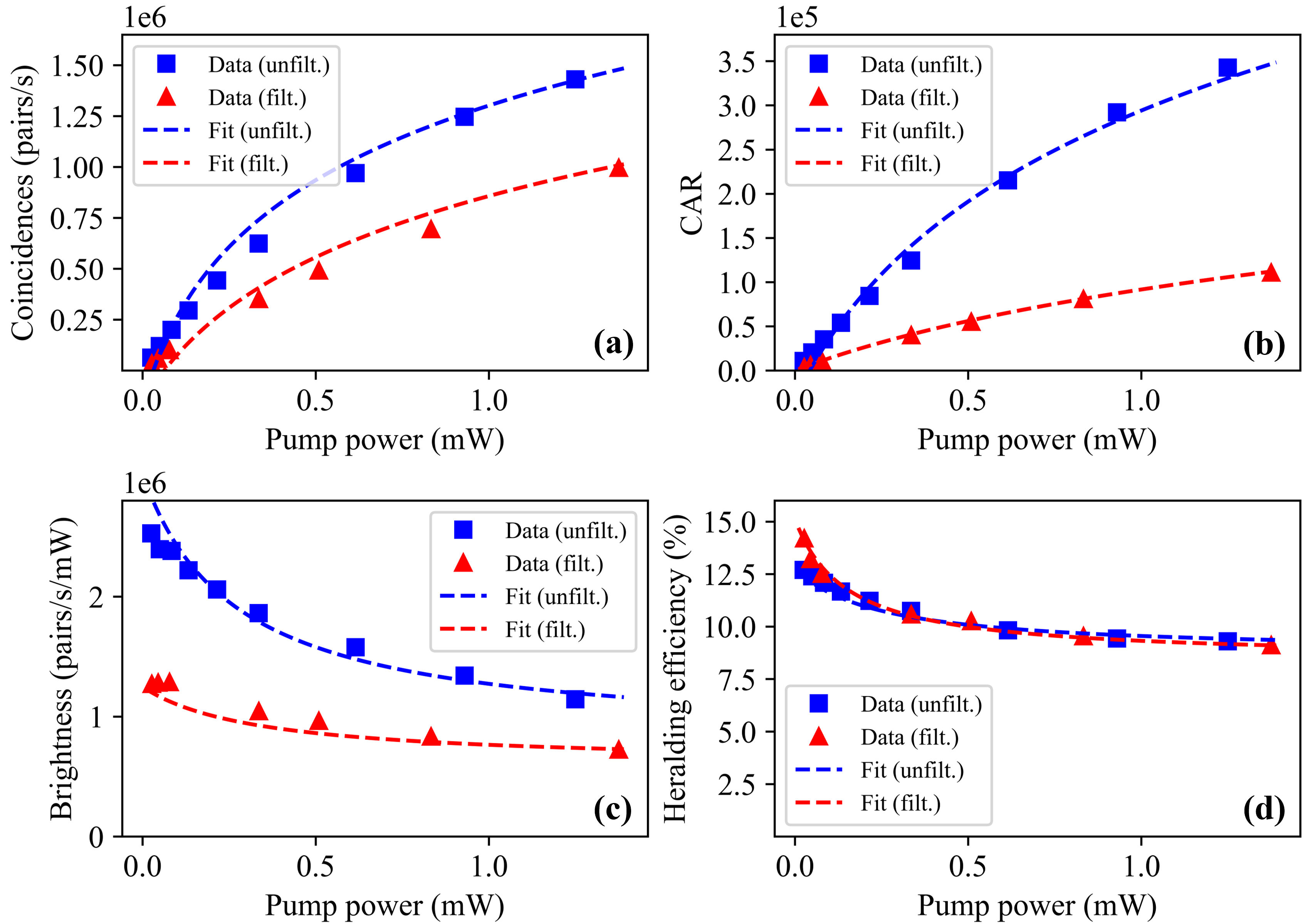}
\caption{Effect of Pump Power. The detected (a) count rates, (b) coincidence-to-accident (CAR) ratio, (c) brightness values and (c) heralding efficiencies achieved by the source, shown against the pump power for the case of no spectral filtering and the addition of a 10~nm bandpass filter centred at 780~nm applied to the signal arm.}
\label{fig:brightness}
\end{figure}

The dependence of the source performance on the pump power was characterized by the detected coincidences as shown in Fig.~\ref{fig:brightness}. The experiment was performed under two different filtering conditions - with and without the bandpass filter of 10~nm FWHM in the signal arm (780~nm). This was limited to pump powers below 1.5~mW to avoid the impact of saturation on the detectors at very high count rates. To keep the pump wavelength constant to maintain the phase matching, the laser current was kept constant but the effective optical power was varied using HWP1 together with the PBS. The same trend is seen in the coincidence-to-accident ratio (CAR) (Fig.~\ref{fig:brightness} (b)), calculated using the coincidence window, $\Delta\tau = 20$~ns. The detected brightness of the source, defined as the coincidences per milliWatt of pump power, decreases as the pump power increases, as shown in Fig.~\ref{fig:brightness}~(c). In the filtered case, this decrease occurs at a much slower rate. Finally, Fig.~\ref{fig:brightness}~(d) shows the heralding efficiency for the two cases. Even with a difference in coincidences, the corresponding efficiencies remain the same for both filtered and unfiltered cases. This shows that the source efficiency is maintained with different filtering conditions. The fits given on each of the plots are derived from the observed trend in the collected data. The trend in the coincidences and the CAR are of the form $a \log((b P )+c)$ and the brightness and heralding efficiency fitting functions are of the form $a + 1/(P + b)$, where $P$ is the pump power and $a$, $b$ and $c$ are constant fitting parameters, unique to each dataset.
\begin{figure}t]
\centering\includegraphics[width=0.85\linewidth]{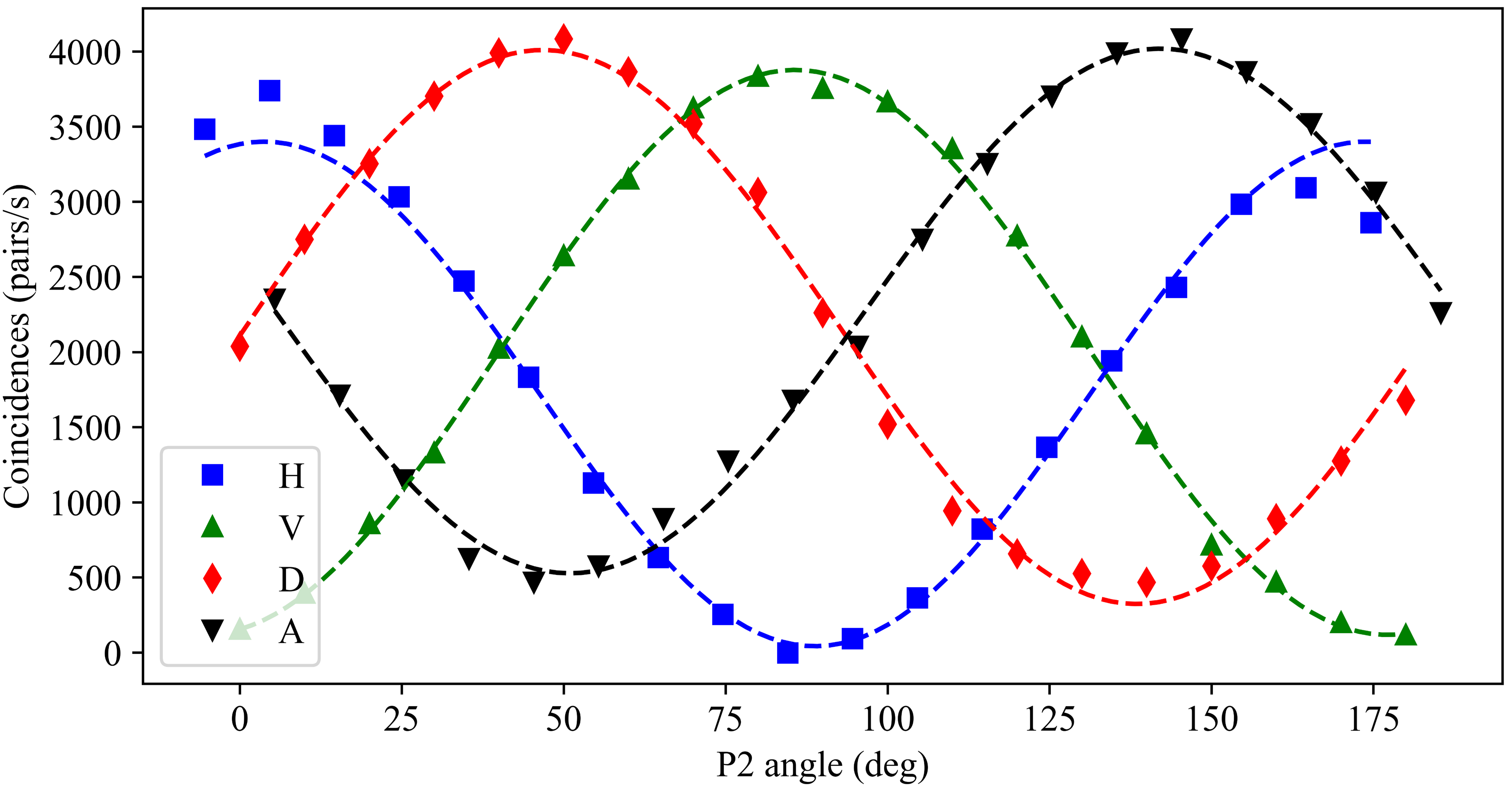}
\caption{Visibility measurements. The coincidence count rate was measured whilst varying the the idler polarizer, P2, angle for fixed signal polarizations H, V, D \& A. The fidelity with the Bell state $|\Phi^+\rangle$ was estimated as 94.1\%$\pm$2.1\%. Data taken without spectral filtering.}
\label{fig:visibility}
\end{figure}

To quantify the entanglement produced by the source, polarization correlation measurements on photon-pairs were carried out using a calibrated polarizer on both the signal (P1) and idler (P2) arms. In this two-linear polarizer setting, the signal photon is projected into one polarization state using a fixed polarizer (P1) and the polarizer on the idler detector arm is swept through 180 degrees. With the angle P1 set to zero (H projection), the coincidences were recorded for different angles P2. Similar measurements were carried out for P1 angles 45$^{\circ}$, 90$^{\circ}$ \& 135$^{\circ}$, which correspond to D, V \& A polarization projections. Fig.~\ref{fig:visibility} shows the plots of variation of the coincidences with the angle P2 for the projections H, V, D \& A projections on P1. From the plots, the visibilities were estimated to be 98.8\%~$\pm$~1.2\% (H), 97.0\%~$\pm$~1.1\% (V), 89.3\%~$\pm$~1.0\% (D) \& 91.2\%~$\pm$~1.0\% (A) and the fidelity with $|\Phi^+\rangle$ is 94.1\%~$\pm$~2.1\%. With this, the quantum bit error rate is estimated to be 5.9\%.

An important feature of the FDLI design is the use of corner cube retroreflectors (CCR), which are insensitive to tip-tilt misalignment~\cite{acebron05}. Such retroreflectors have applications in the construction of high-precision stable optical systems that involve interferometers~\cite{zamiela13}. In this source, the insensitivity of CCR to tip-tilt misalignment is verified by monitoring the singles and pair rates with angular misalignment as well as any displacement misalignment from the mounting of the CCR, though this is minimized by the use of gimbal mount. The experiment was carried out for two different cases - with and without compensating the fiber coupling due to any unavoidable lateral misalignment. 

\begin{figure}[b]
\centering\includegraphics[width=0.85\linewidth]{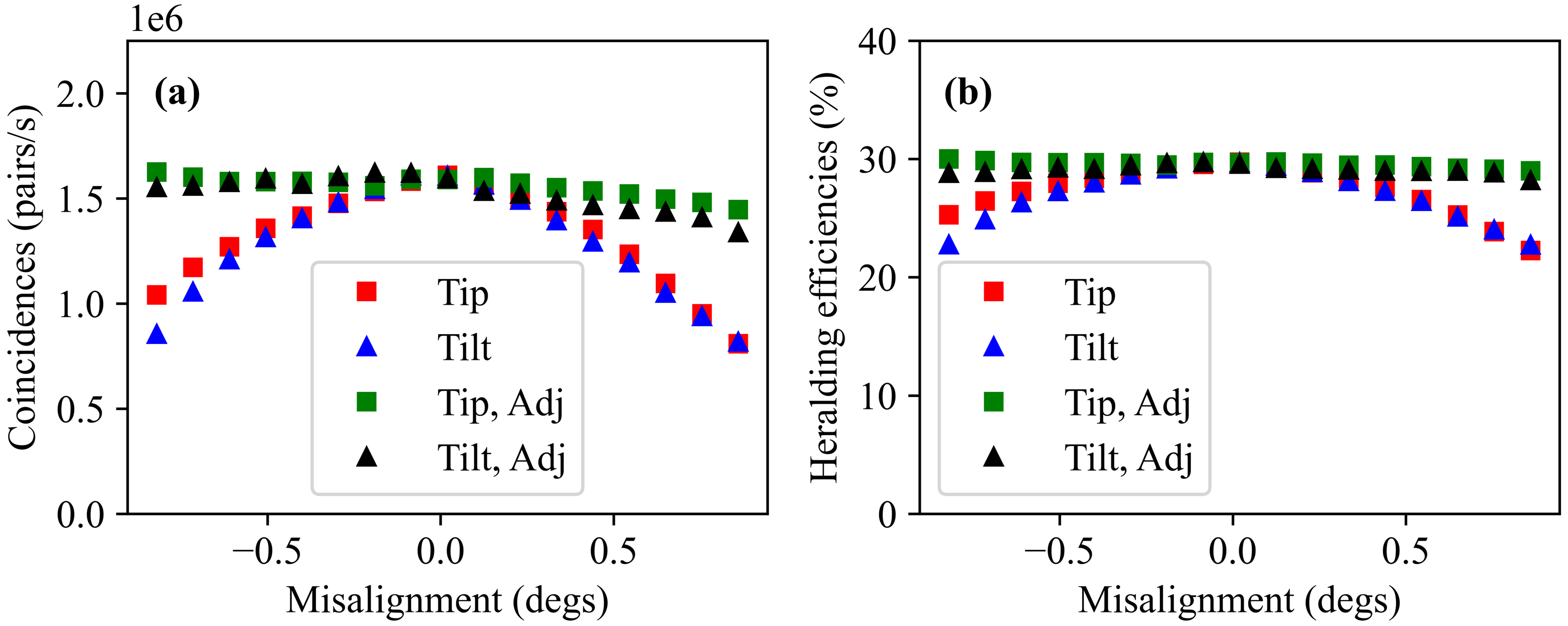}
\caption{Misalignment Robustness. The stability of the interferometer alignment when introducing CCR tip and tilt misalignment is shown. For the tip and tilt misalignment for the adjusted and non-adjusted cases, (a) shows the effect on the coincidences and (b) shows the impact on the heralding efficiency.}
\label{fig:align}
\end{figure}

The results for both cases are given in Fig.~\ref{fig:align}. Without compensation, the counts and heralding efficiencies degrade to approximately 1 M pairs/s and 20\%, which would still be suitable for many applications. However, this degradation can be corrected with slight adjustments in the fiber coupling unit, as shown in Fig. \ref{fig:align}~(a) and (b), where the coincidences and efficiencies are well maintained. In the case where slight adjustments were made to the fiber coupling, the performance is maintained above 80\% of the optimal position and above 95\% for the introduction of a -0.85$^\circ$ shift, seen on the left-hand side of the figure. 
Beam steering systems, such as the piezo steering mirrors used in source in the photon pair source on the Micius satellite~\cite{Yin2017}, can be used to maintain coupling in deployed systems. However, this may not be required for this system as high performance is maintained the expected range a system would experience when deployed.
But with or without active alignment, this configuration has demonstrated stability against tip-tilt misalignment in the CCR, used to fold the interferometer, gaining the double-pass and reduced footprint. 

\begin{figure}[b]
\centering\includegraphics[width=1\linewidth]{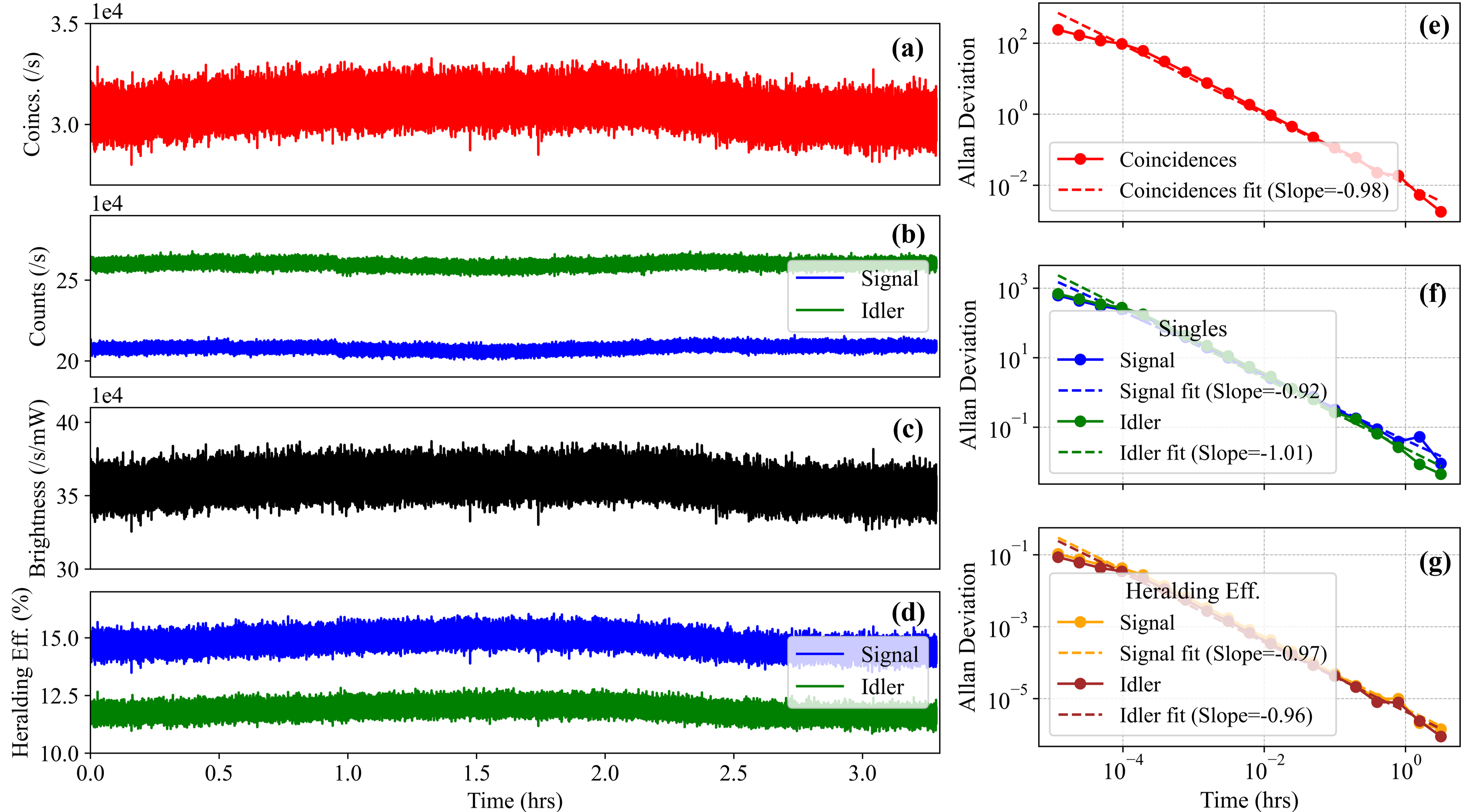}
\caption{Performance stability. Over 3.3 hrs the performance of the source was monitored under constant external conditions. (a) - (d) show the monitoring of the coincidences, single counts, brightness and the heralding efficiencies over 3.3 hours. (e) - (g) show the allan deviation of the coincidences, singles and heralding efficiency. The laser diode driving current was 40~mA, giving an average pump power of 86 $\mu$W after attenuation. }
\label{fig:monitor}
\end{figure}

For characterization of the optical stability, the source was operated under a constant background condition for several hours without changing the source and detection settings. Fig.~\ref{fig:monitor}~(a) - (d) show the coincidences, single counts, brightnesses and heralding efficiencies monitored over the operational time. The long-term stability of the source can be estimated from the Allan deviation. The Allan deviation of a given measured parameter $P$ with respect to the integration time $T$ is defined as~\cite{allan1966statistics},
\begin{equation}\label{eqn:allan}
    \sigma(T=N\tau)=\sqrt{\frac{1}{2}\left\langle \Delta P^2\right\rangle}=\sqrt{\frac{1}{2(N_{\text{tot}}-N)}\sum_{i=1}^{N_{\text{tot}}-N}(P_{i+N}-P)^2}.
\end{equation}
Here, the overall integration time, $T=N\tau$, is divided into $N$ samples with time interval $\tau=13$~ms and the total number of samples $N_{\text{tot}}=$ 907083. Fig. \ref{fig:monitor} (e) - (g) show the plots in logarithmic scale of allan deviation with pair counts, singles counts and heralding efficiencies for the overall monitor time. In all these, the linear fitting of that data gives a negative slope close to -1, which suggests that averaging over longer periods reduces the noise, implying that fluctuations are likely due to random white noise rather than systematic drifts. This means the system does not exhibit significant drift over the time period considered.

\section{Discussion and Conclusion}

The folded geometry of the source not only reduces the number of components required, allowing a smaller footprint, but also provides a double-pass of the pump through the crystal and thereby enhancing the pair production rate. Table~\ref{tab:perf} summarizes the performance values of the source and compares them to those of similar designs. The source in this work has, to the best of our knowledge, demonstrated a brightness value highest among those reported for a broadband pump, but with moderate heralding efficiencies. The reported heralding efficiency was recorded simultaneously with the high brightness values, however it can be seen in Figs.~\ref{fig:phasematching}~(b) and \ref{fig:align}~(b) that heralding efficiencies up to 30\% are achievable. The source reported in~\cite{Steinlechner_2013} also utilizes the advantage of a double pass. However, in that work, the propagation through the achromatic waveplate after the SPDC crystal may reduce the pump power available for the down-conversion during second pass which gives an unbalance entangled state. In our source, introducing any additional optics between the PPKTP and the CCR will not affect the balancing of HH and VV pairs as both the pump parts through the crystal will have equal effect.

As discussed previously, there is a trade-off between brightness and heralding efficiency, the result in Table~\ref{tab:perf} is where the brightness has been optimized. As our pump focus is designed to be maintained over both passes through the PPKTP as well as the during the reflection, a total distance of roughly 3.5~cm, this increases our focal parameter, potentially at the expense of high heralding efficiencies. Reducing the space between the PPKTP and CCR could minimize this effect. Further improvements in the engineering of the pump, such as using a narrower linewidth laser, and detection systems, such as alternative pair separation methods, could also allow better detected brightnesses and heralding efficiencies to be achieved simultaneously. 

\begin{table}[t]
    \caption{Summary of the performance parameters. We compare the presented source with other reported polarization entangled sources for simultaneously achieved brightness and heralding efficiency. $^\dagger$Maximum achieved heralding efficiency.}
    \label{tab:perf}
    \centering
    \renewcommand{\arraystretch}{1.2} 
    \begin{tabular}{
        >{\raggedright}p{2cm}  
        >{\raggedright}p{3cm}  
        >{\centering\arraybackslash}p{2cm}  
        >{\centering\arraybackslash}p{2cm}  
        >{\centering\arraybackslash}p{2cm}  
    } 
        \toprule
        \textbf{Source design} & \textit{Configuration} & \textit{Wavelengths (nm)} & \textit{Brightness (Mpairs/s/mW)} & \textit{Heralding efficiency (\%)}\\ 
        \midrule
        Sagnac~\cite{steinlechner2014} & Loop, double pass & 783, 839 & 1.0 & 40\\
        LDI~\cite{Lohrmann2020} & Linear, single pass, single displacement & 810, 810 & 0.56 & 21\\
        Double LDI~\cite{fazili2024simple} & Linear, single pass, double displacement & 785, 841 & 1.9  & 32\\
        Folded sandwich~\cite{Steinlechner_2013} & Linear, back-reflected double pass & 783, 839 & 1.1 & 16\\
        This work (Folded LDI) & Linear, retro-reflected double pass, single displacement & 780, 842 & 2.5 & 14 (30$^\dagger$)\\
        \bottomrule
    \end{tabular}
\end{table}

A source of polarization-entangled photon-pairs based on a folded linear displacement interferometer has been demonstrated, showing multiple benefits compared to similar linear designs. Folding the interferometer reduces the source footprint and allows for single beam displacing and polarization-flipping components to be repurposed, whilst enhancing the source performance with a double-pass configuration. The insensitivity of the cornercube retroreflector to angular misalignment was utilized to maintain optical stability of the source. The source has a good Bell state fidelity and has demonstrated stability in performance over multiple hours.  These qualities: reduced footprint, high pair-production rates, and assured stability, all directly lend themselves to satellite deployment, and thus, towards the realization of space-enabled quantum communication networks.

\begin{backmatter}

\bmsection{Funding}
Innovate UK project NextSTEPS 10031944, EPSRC Research Excellence Award Studentship (EP/T517938/1), EPSRC Quantum Technology Hub in Quantum Communication (EP/T001011/1), EPSRC International Network in Space Quantum Technologies (EP/W027011/1), and the EPSRC Integrated Quantum Networks Hub (EP/Z533208/1).

\bmsection{Acknowledgments}
We acknowledge discussions with Craft Prospect and Alter Technology relating to this work.

\bmsection{Disclosures}
The authors declare no conflicts of interest.

\bmsection{Data Availability Statement}
Data available upon request.

\end{backmatter}

\newpage

\bibliography{sample}

\begin{thebibliography}{10}
\newcommand{\enquote}[1]{``#1''}

\bibitem{Pirandola_2018}
S.~Pirandola, B.~R. Bardhan, T.~Gehring, \emph{et~al.}, \enquote{Advances in photonic quantum sensing,} {\protect\JournalTitle{Nature Photonics}} \textbf{12}, 724–733 (2018).

\bibitem{camphausen23}
R.~Camphausen, A.~Perna, A.~Cuevas, \emph{et~al.}, \enquote{Fast quantum-enhanced imaging with visible-wavelength entangled photons,} {\protect\JournalTitle{Optics Express}} \textbf{31}, 6039--6050 (2023).

\bibitem{ekert91}
A.~K. Ekert, \enquote{Quantum cryptography based on {B}ell's theorem,} {\protect\JournalTitle{Phys. Rev. Lett.}} \textbf{67}, 661--663 (1991).

\bibitem{BBM92}
C.~H. Bennett, G.~Brassard, and N.~D. Mermin, \enquote{Quantum cryptography without {B}ell's theorem,} {\protect\JournalTitle{Phys. Rev. Lett.}} \textbf{68}, 557--559 (1992).

\bibitem{Caleffi_2024}
M.~Caleffi, M.~Amoretti, D.~Ferrari, \emph{et~al.}, \enquote{Distributed quantum computing: A survey,} {\protect\JournalTitle{Computer Networks}} \textbf{254}, 110672 (2024).

\bibitem{pant2017}
M.~Pant, H.~Krovi, D.~Towsley, \emph{et~al.}, \enquote{Routing entanglement in the quantum internet,} {\protect\JournalTitle{npj Quantum Information}} \textbf{5}, 25 (2017).

\bibitem{Liao_2017}
S.-K. Liao, W.-Q. Cai, W.-Y. Liu, \emph{et~al.}, \enquote{Satellite-to-ground quantum key distribution,} {\protect\JournalTitle{Nature}} \textbf{549}, 43–47 (2017).

\bibitem{oi2017}
D.~K.~L. Oi, A.~Ling, J.~A. Grieve, \emph{et~al.}, \enquote{Nanosatellites for quantum science and technology,} {\protect\JournalTitle{Contemporary Physics}} \textbf{58}, 25--52 (2017).

\bibitem{Oi_apr2017}
D.~K. Oi, A.~Ling, G.~Vallone, \emph{et~al.}, \enquote{{CubeSat} quantum communications mission,} {\protect\JournalTitle{EPJ Quantum Technology}} \textbf{4}, 1--20 (2017).

\bibitem{Yin2017}
J.~Yin, Y.~Cao, Y.-H. Li, \emph{et~al.}, \enquote{Satellite-based entanglement distribution over 1200 kilometers,} {\protect\JournalTitle{Science}} \textbf{356}, 1140--1144 (2017).

\bibitem{Lu2022}
C.-Y. Lu, Y.~Cao, C.-Z. Peng, and J.-W. Pan, \enquote{Micius quantum experiments in space,} {\protect\JournalTitle{Rev. Mod. Phys.}} \textbf{94}, 035001 (2022).

\bibitem{villar2020}
A.~Villar, A.~Lohrmann, X.~Bai, \emph{et~al.}, \enquote{Entanglement demonstration on board a nano-satellite,} {\protect\JournalTitle{Optica}} \textbf{7}, 734--737 (2020).

\bibitem{schimpf21}
C.~Schimpf, M.~Reindl, F.~Basset, \emph{et~al.}, \enquote{Quantum dots as potential sources of strongly entangled photons: Perspectives and challenges for applications in quantum networks,} {\protect\JournalTitle{Applied Physics Letters}} \textbf{118}, 100502 (2021).

\bibitem{wang24}
H.~Wang, Q.~Zeng, H.~Ma, and Z.~Yuan, \enquote{Progress on chip-based spontaneous four-wave mixing quantum light sources,} {\protect\JournalTitle{Advanced Devices \& Instrumentation}} \textbf{5} (2024).

\bibitem{Li_2015}
Y.~Li, Z.-Y. Zhou, D.-S. Ding, and B.-S. Shi, \enquote{{CW}-pumped telecom band polarization entangled photon pair generation in a sagnac interferometer,} {\protect\JournalTitle{Optics Express}} \textbf{23}, 28792 (2015).

\bibitem{Olislager2009}
L.~Olislager, J.~Cussey, A.-T. Nguyen, \emph{et~al.}, \enquote{Frequency bin entangled photons,} {\protect\JournalTitle{Physical Review A}} \textbf{82}, 013804 (2009).

\bibitem{Kwon13}
O.~Kwon, K.-K. Park, Y.-S. Ra, \emph{et~al.}, \enquote{Time-bin entangled photon pairs from spontaneous parametric down-conversion pumped by a {CW} multi-mode diode laser,} {\protect\JournalTitle{Opt. Express}} \textbf{21}, 25492--25500 (2013).

\bibitem{Anwar2021}
A.~Anwar, C.~Perumangatt, F.~Steinlechner, \emph{et~al.}, \enquote{Entangled photon-pair sources based on three-wave mixing in bulk crystals,} {\protect\JournalTitle{Review of Scientific Instruments}} \textbf{92}, 041101 (2021).

\bibitem{steinlechner2014}
F.~Steinlechner, M.~Gilaberte~Basset, M.~Jofre, \emph{et~al.}, \enquote{Efficient heralding of polarization-entangled photons from type-0 and type-ii spontaneous parametric downconversion in periodically poled ktiopo4,} {\protect\JournalTitle{Journal of the Optical Society of America B}} \textbf{31}, 2068--2076 (2014).

\bibitem{Anwar22}
A.~Anwar, C.~Perumangatt, A.~Villar, \emph{et~al.}, \enquote{Development of compact entangled photon-pair sources for satellites,} {\protect\JournalTitle{Applied Physics Letters}} \textbf{121}, 220503 (2022).

\bibitem{villar2018}
A.~Villar, A.~Lohrmann, and A.~Ling, \enquote{Experimental entangled photon pair generation using crystals with parallel optical axes,} {\protect\JournalTitle{Optics Express}} \textbf{26}, 12396--12402 (2018).

\bibitem{perumangatt2021realizing}
C.~Perumangatt, T.~Vergoossen, A.~Lohrmann, \emph{et~al.}, \enquote{Realizing quantum nodes in space for cost-effective, global quantum communication: in-orbit results and next steps,} in \emph{Quantum Computing, Communication, and Simulation,}  vol. 11699 (SPIE, 2021), p. 1169904.

\bibitem{Lohrmann2020}
A.~Lohrmann, C.~Perumangatt, A.~Villar, and A.~Ling, \enquote{Broadband pumped polarization entangled photon-pair source in a linear beam displacement interferometer,} {\protect\JournalTitle{Applied Physics Letters}} \textbf{116}, 021101 (2020).

\bibitem{horn2019auto}
R.~Horn and T.~Jennewein, \enquote{Auto-balancing and robust interferometer designs for polarization entangled photon sources,} {\protect\JournalTitle{Optics Express}} \textbf{27}, 17369--17376 (2019).

\bibitem{fazili2024simple}
R.~Fazili, P.~S. Chauhan, U.~Chandrashekara, \emph{et~al.}, \enquote{Simple but efficient polarization-entangled photon sources,} {\protect\JournalTitle{Journal of the Optical Society of America B}} \textbf{41}, 2692--2701 (2024).

\bibitem{Liu1997}
J.~Liu and R.~Azzam, \enquote{Polarization properties of corner-cube retroreflectors: Theory and experiment,} {\protect\JournalTitle{Applied Optics}} \textbf{36}, 1553--9 (1997).

\bibitem{Li2024}
N.~Li, F.~Zheng, Q.~Feng, \emph{et~al.}, \enquote{Analysis and experimental verification of the polarization characteristics of cube-corner reflectors,} {\protect\JournalTitle{Applied Optics}} \textbf{63}, 3462--3469 (2024).

\bibitem{BOYD202065}
R.~W. Boyd, \enquote{Chapter 2 - wave-equation description of nonlinear optical interactions,} in \emph{Nonlinear Optics (Fourth Edition),}  R.~W. Boyd, ed. (Academic Press, 2020), pp. 65--135, fourth edition ed.

\bibitem{dixon14}
P.~B. Dixon, D.~Rosenberg, V.~Stelmakh, \emph{et~al.}, \enquote{Heralding efficiency and correlated-mode coupling of near-{IR} fiber-coupled photon pairs,} {\protect\JournalTitle{Phys. Rev. A}} \textbf{90}, 043804 (2014).

\bibitem{acebron05}
J.~Acebron and R.~Spigler, \enquote{The magic mirror property of the cube corner,} {\protect\JournalTitle{Mathematics Magazine}} \textbf{78}, 308--311 (2005).

\bibitem{zamiela13}
G.~Zamiela and M.~Dobosz, \enquote{Corner cube reflector lateral displacement evaluation simultaneously with interferometer length measurement,} {\protect\JournalTitle{Optics \& Laser Technology}} \textbf{50}, 118--124 (2013).

\bibitem{allan1966statistics}
D.~W. Allan, \enquote{Statistics of atomic frequency standards,} {\protect\JournalTitle{Proceedings of the IEEE}} \textbf{54}, 221--230 (1966).

\bibitem{Steinlechner_2013}
F.~Steinlechner, S.~Ramelow, M.~Jofre, \emph{et~al.}, \enquote{Phase-stable source of polarization-entangled photons in a linear double-pass configuration,} {\protect\JournalTitle{Optics Express}} \textbf{21}, 11943 (2013).

\end{thebibliography}

\appendix

\section{Polarization characterization of CCRs} \label{appendix}

\begin{figure}[ht!]
\centering\includegraphics[width=1\linewidth]{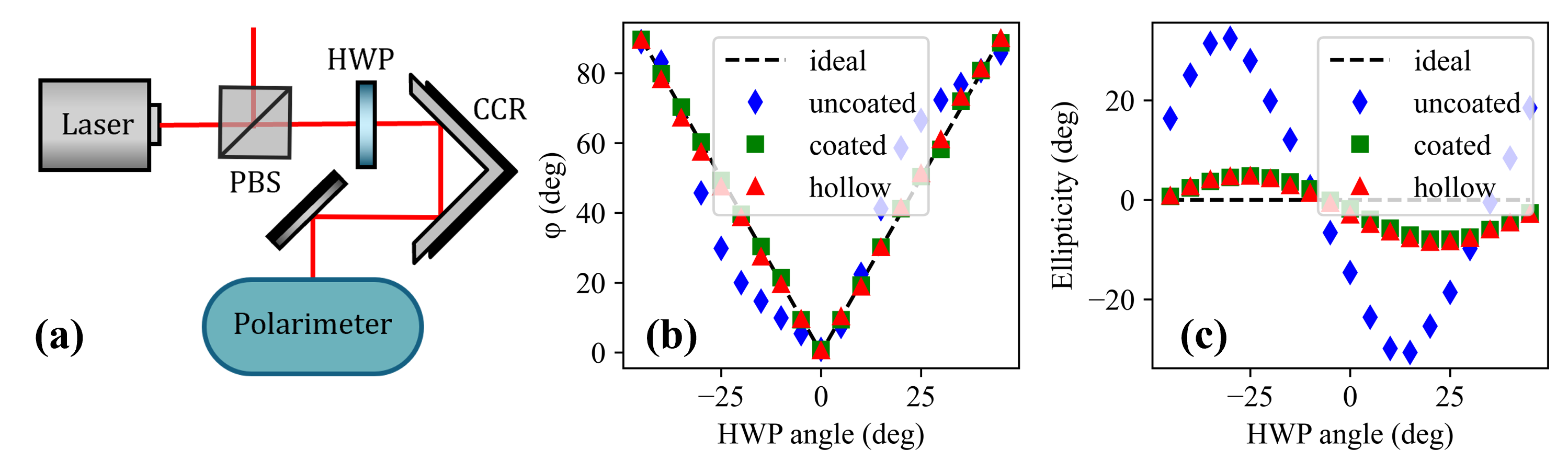}
\caption{Experimental set up, in (a), for the characterization of the CCRs, conducted by varying the polarization state incident on the CCR with a linearly polarized laser, PBS and HWP, and analyzing the reflected polarization state with a polarimeter. For the HWP angle, (b) angle $\varphi$ and (c) ellipticity are given, compared to the ideal case. }
\label{fig:RR}
\end{figure}

The polarization of light can be altered by reflection, particularly from dielectric surfaces. As CCRs rely on multiple reflections off orthogonal surfaces, if the s- and p-polarization components experience different amounts of reflectance or phase shift, this will produce a change in polarization of the outgoing beam. This is observed in some CCRs, for example reflecting linearly polarized light as circular. Thus, for the generation of polarization-entangled photon-pairs it was vital to minimize, or account for, any polarization changes caused by reflection on the CCR. The use of metallic coatings has been found to limit the effect of induced ellipticity and also that hollow CCRs did not experience the same issues\cite{Li2024}. Therefore, in the development of the source, three CCRs were characterized; an uncoated solid CCR (Thorlabs, PS974M), a gold-coated solid CCR (Thorlabs, PS977M-M01B), and a silver-coated hollow CCR (Thorlabs, HRR202-P01). The experimental set-up for characterizing the CCRs is shown in Fig. \ref{fig:RR} (a). A CW laser of wavelength 785~nm (Edinburgh Instruments, EPL-785) is incident on a CCR. The retroreflected light is detected by the polarimeter (Sch\"after and Kirchhoff, SK010PA). The polarization of the incident beam is controlled by the combination of a PBS and HWP. Figs. \ref{fig:RR} (b) and (c), show the orientation of the output polarization state, $\varphi$, and the degree of ellipticity against the angle of the HWP, respectively.

In Fig. \ref{fig:RR} (b), the hollow and gold-coated CCRs can be seen to match the ideal state well over the whole range, whereas the uncoated CCR shows significant deviation in the state orientation, though it recovers at angles -45$^\circ$, 0$^\circ$ and 45$^\circ$, for the H and V polarizations. As shown in Fig. \ref{fig:RR} (c), the reflected light from the uncoated solid CCR has a varying ellipticity with different linear polarization inputs. Therefore, uncoated solid CCRs are not appropriate for polarization-sensitive optical systems, like polarization-entangled photon-pair sources. Although some ellipticity is induced the retroreflected state from the coated and hollow CRRs, the effect is much smaller and the state is consistent at H and V polarizations required for implementing the source. 

Although the gold-coated solid and the hollow silver-coated CCRs behave similarly in this polarization characterization, the gold coating has drastically reduced reflectivity ($< 50\%$) at the pump wavelength, reducing any benefit gained from the double-pass in this source design. Therefore, the silver-coated hollow CCR was used to fold the interferometer for the source described above.

\end{document}